\documentstyle[aps,multicol,prl,epsf]{revtex}
\newcommand{\etal}{{\it et al.}}

\def\gtorder{\mathrel{\raise.3ex\hbox{$>$}\mkern-14mu
             \lower0.6ex\hbox{$\sim$}}}
\def\ltorder{\mathrel{\raise.3ex\hbox{$<$}\mkern-14mu
             \lower0.6ex\hbox{$\sim$}}}

\def\deg{^\circ}

\begin{document}

\draft 
\twocolumn[\hsize\textwidth\columnwidth\hsize\csname@twocolumnfalse%
\endcsname

\title{$x$- and $\xi$-scaling of the Nuclear Structure Function at Large $x$.}

\author{J. Arrington$^{3,1}$, 
C. S. Armstrong$^{12}$, 
T. Averett$^{3,12}$, 
O. K. Baker$^{5,10}$, 
L. de Bever$^{2}$, 
C. W. Bochna$^{6}$, 
W. Boeglin$^{4}$, 
B. Bray$^{3}$, 
R. D. Carlini$^{10}$, 
G. Collins$^{7}$, 
C. Cothran$^{11}$, 
D. Crabb$^{11}$, 
D. Day$^{11}$, 
J. A. Dunne$^{10}$, 
D. Dutta$^{8}$, 
R. Ent$^{10}$, 
B. W. Filippone$^{3}$, 
A. Honegger$^{2}$, 
E. W. Hughes$^{3}$, 
J. Jensen$^{3}$, 
J. Jourdan$^{2}$, 
C. E. Keppel$^{5,10}$, 
D. M. Koltenuk$^{9}$, 
R. Lindgren$^{11}$, 
A. Lung$^{7}$, 
D. J Mack$^{10}$, 
J. McCarthy$^{11}$, 
R. D. McKeown$^{3}$, 
D. Meekins$^{12}$, 
J. H. Mitchell$^{10}$, 
H. G. Mkrtchyan$^{10}$, 
G. Niculescu$^{5,14}$, 
I. Niculescu$^{5,13}$, 
T. Petitjean$^{2}$, 
O. Rondon$^{11}$, 
I. Sick$^{2}$, 
C. Smith$^{11}$, 
B. Terburg$^{6}$  
W. F. Vulcan$^{10}$, 
S. A. Wood$^{10}$, 
C. Yan$^{10}$, 
J. Zhao$^{2}$, 
B. Zihlmann$^{11}$ }

\address {$^{1}$Argonne National Lab, Argonne, IL 60439\\
$^{2}$University of Basel, Basel Switzerland\\
$^{3}$Kellogg Radiation Laboratory, California Institute of Technology, Pasadena CA 91125\\
$^{4}$Florida International University, University Park FL 33199\\
$^{5}$Hampton University, Hampton VA 23668\\
$^{6}$University of Illinois, Urbana-Champaign IL 61801\\
$^{7}$University of Maryland, College Park MD 20742\\
$^{8}$Northwestern University, Evanston IL 60201\\
$^{9}$University of Pennsylvania, Philadelphia PA 19104\\
$^{10}$Thomas Jefferson National Accelerator Facility, Newport News VA 23606\\
$^{11}$University of Virginia, Charlottesville VA 22901\\
$^{12}$College of William and Mary, Williamsburg, VA 23187\\
$^{13}$George Washington University, Washington, D.C. 20052\\
$^{14}$Ohio University, Athens, Ohio, 45701\\
} 

\date{September 14, 2000}
\maketitle

\begin{abstract}
Inclusive electron scattering data are presented for $^2$H and Fe
targets at an incident electron energy of 4.045 GeV for a range of momentum
transfers from $Q^2 = 1$ to 7 (GeV/c)$^2$.  Data were taken at Jefferson
Laboratory for low values of energy loss, corresponding to values of Bjorken
$x \gtorder 1$. The structure functions do not show scaling in $x$ in this
range, where inelastic scattering is not expected to dominate the cross section. 
The data do show scaling, however, in the Nachtmann variable $\xi$.  This
scaling may be the result of Bloom Gilman duality in the nucleon
structure function combined with the Fermi motion of the nucleons in the
nucleus.  The resulting extension of scaling to larger values of $\xi$ opens
up the possibility of accessing nuclear structure functions in the high-$x$ region at
lower values of $Q^2$ than previously believed.

\medskip
\end{abstract}
]

Deep inelastic electron scattering (DIS) from protons has provided a
wealth of information on the parton structure of the nucleon.  In general, the
nucleon structure functions $W_1$ and $W_2$ depend on both the energy transfer
($\nu$) and the square of the four-momentum transfer (-$Q^2$).  In the Bjorken
limit of infinite momentum and energy transfer, the structure functions
depend only on the ratio of $Q^2 / \nu$ (modulo QCD scaling violations). Thus, when taken as a function of
Bjorken $x$ (= $Q^2/2M\nu$, where $M$ is the mass of the proton), the
structure functions are independent of $Q^2$.  In the parton model, $x$ is
interpreted as the longitudinal momentum fraction of the struck quark, and the
structure function can be related to the quark momentum distributions. 
This scaling was observed in high energy electron-proton scattering at SLAC,
confirming the parton picture of the nucleon. Violations of Bjorken scaling
arise at low $Q^2$ due to effects coming from kinematic corrections and
higher-twist effects.  A better scaling variable for finite $Q^2$ comes from
the operator product expansion treatment of DIS, as was shown in \cite{georgi}.
Using the Nachtmann variable $\xi = 2x/(1+\sqrt{1+4M^2x^2/Q^2})$ 
avoids additional scaling violations arising from finite $Q^2$ corrections to
$x$-scaling (which is derived in the infinite momentum limit).

Scaling in $x$ should also be seen in electron-nucleus scattering as both
$\nu$ and $Q^2$ approach $\infty$.  Because $x$ represents a momentum fraction,
it must be between 0 and 1 for scattering from a nucleon.  When scattering
from a nucleus, $x$ can vary between 0 and $A$, the number of nucleons,
due to the nucleon momentum sharing.  At
finite $Q^2$ and large $x$ ($x \gtorder 1$), additional scaling violations
come from quasielastic (QE) scattering off of a nucleon in the nucleus, rather
than scattering off of a single quasi-free quark.  The quasielastic contribution
to the cross section decreases with respect to the inelastic contributions as
$Q^2$ increases due to the nucleon elastic form factor, but QE scattering
dominates at very low energy loss (corresponding to $x$$>$1) up to large
values of $Q^2$.

Previous measurements  of inclusive electron scattering from nuclei for  $x
\ltorder 3$ and $Q^2 \ltorder 3$ (GeV/c)$^2$ (SLAC experiment NE3
\cite{ne3_x}) showed scaling for $x \leq 0.4$, but a significant $Q^2$
dependence for larger $x$ values.  For these $x$ values, the momentum transfer
is low enough that quasielastic and resonance contributions to the scattering
violate the expected scaling in $x$.  When the structure function was
examined as a function of $\xi$, the behavior was completely different.  The
data appeared to be approaching a universal curve as $Q^2$ increased, even in
regions where the scattering was predominantly quasielastic. It was noted
that this behavior is similar to the local duality observed
by Bloom and Gilman \cite{bgdual,derujula} in the proton structure function.
Local duality is essentially the observation that the
structure function in the resonance region has the same behavior as the deep
inelastic structure function, when averaged over a range in $\xi$. It was
suggested \cite{ne3_x} that in the nucleus, the nucleon momentum distribution could
effectively perform this averaging of the structure function, causing the QE
and DIS contributions to have the same $Q^2$ behavior.  Later it was suggested
\cite{accidental} that the apparent scaling might instead come from an
accidental cancellation of $Q^2$ dependent terms, and would occur only for a
limited range of momentum transfers (up to $Q^2 \sim 7.0$ (GeV/c)$^2$).  More
recent measurements (SLAC experiment NE18 \cite{ne18}) showed continued
scaling behavior up to $Q^2 = 6.8$ (GeV/c)$^2$, but the data were limited to
values of $x$ very close to 1.

The present data, from experiment E89-008 at Jefferson Lab, were taken with an
electron beam energy of 4.045 GeV for scattering angles between 15 and 74
degrees, covering a $Q^2$ range from 1 to 7 (GeV/c)$^2$. The scattered
electrons were measured in the High Momentum Spectrometer (HMS) and Short
Orbit Spectrometer (SOS) in Hall C. Data were taken using cryogenic hydrogen
and deuterium targets and solid targets of C, Fe, and Au.  Details of the
experiment and cross section extraction can be found in refs. \cite{89008_y}
and \cite{jra}.

For unpolarized scattering from a nucleus, the inclusive cross section (in the
one-photon-exchange approximation) can be written as:

\begin{equation}
\label{results_x1}
\frac{d\sigma}{d\Omega dE^\prime}=\sigma_{Mott}
\left[ W_2 + 2W_1 \tan^2(\theta /2)\right],
\end{equation}
where $\sigma_{Mott}=4\alpha^2E^2\cos^2(\theta/2)/Q^4$, $\theta$ is the
scattering angle, and $W_1(\nu,Q^2)$, $W_2(\nu,Q^2)$ are the structure
functions. An explicit separation of $W_1$ and $W_2$ requires performing a
Rosenbluth separation, which involves measuring the cross section at a fixed
$\nu$ and $Q^2$ while varying the incident energy and scattering angle.
Because the data is taken at fixed beam energies, we make an assumption about
the ratio of the longitudinal to transverse cross section, $R = \sigma_L /
\sigma_T = (1+\nu^2/Q^2)W_2/W_1-1$, to extract $W_2$. Given a value for $R$, we
can determine the dimensionless structure function $\nu W_2$ directly
from the cross section:

\begin{equation}
\label{results_x2}
\nu W_2=\frac{\nu}{1+\beta} \cdot 
\frac{\frac{d\sigma}{d\Omega dE^\prime}}{\sigma_{Mott}},
\end{equation}
where

\begin{equation}
\label{results_x3}
\beta = 2\tan^2(\theta/2)\frac{1+\frac{\nu^2}{Q^2}}{1+R}.
\end{equation}

For our analysis, we use the parameterization $R = 0.32/Q^2$ \cite{bosted}, and
assign a 100\% uncertainty to this value.  This parameterization comes from the
non-relativistic plane-wave impulse approximation (PWIA) for quasielastic
scattering. It is also consistent with data taken in the DIS region
(0.2$<x<$0.5 for $Q^2$ up to 5 (GeV/c)$^2$) \cite{dasu1,dasu2,dasu3} and a
measurement of $R$ near $x$=1 in a $Q^2$ range similar to that of the present
experiment \cite{bosted}.

For the HMS ($\theta \leq 55\deg$), the systematic uncertainty in the cross
section is typically 3.5-4.5\%, dominated by acceptance, radiative
corrections, and bin centering.  For the high $x$ points, the systematic
uncertainties become larger because of the strong kinematic dependence of the
cross section, but are always smaller than the statistical
uncertainties. The uncertainty in $R$ causes an additional uncertainty in the
extracted structure function of 0.5-5.0\%, which is largest for the largest
scattering angles.  For the SOS ($\theta=74\deg$), the total systematic
uncertainty in the structure function is typically $\sim$12\% (due mostly to
large background from pair production), somewhat larger at the highest values
of $x$.

\begin{figure}[xfe]
\epsfxsize 8.5 cm \epsfysize 6 cm \epsfbox{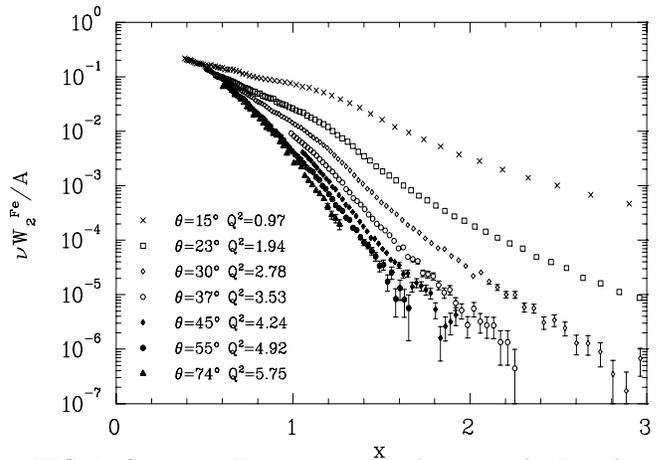}
\caption{Structure Function per nucleon vs $x$ for Iron from the present
measurement. The $Q^2$ values given are for $x = 1$.  Errors shown are
statistical only.}
\label{xfe}
\end{figure}

\begin{figure}[xife]
\epsfxsize 8.5 cm \epsfysize 6 cm \epsfbox{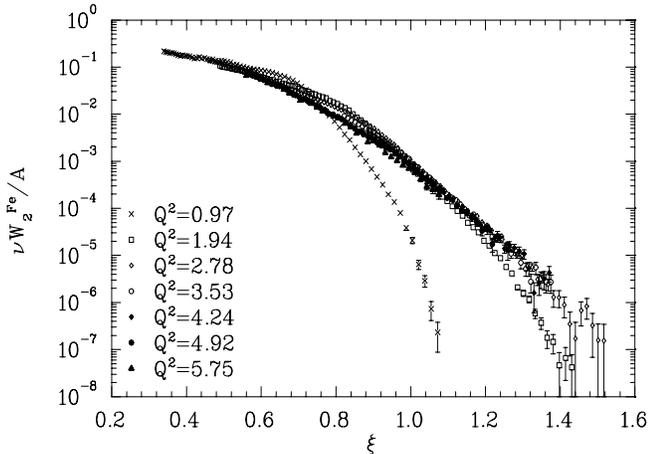}
\caption{Structure Function per nucleon vs $\xi$ for Iron. The $Q^2$ values are given for 
$x$=1.  Errors shown are statistical only.}
\label{xife}
\end{figure}

Figure \ref{xfe} shows the extracted structure function for iron as a function of $x$. 
As in the previous data \cite{ne3_x}, scaling is seen only for values of $x$ significantly
below one, where DIS dominates and resonance and QE contributions are
negligible.  However, when taken as a function of $\xi$ (Fig. \ref{xife}), the
structure function shows scaling for nearly all values of $\xi$.  At low
$\xi$, DIS dominates, and scaling behavior is expected from the parton
model.  For intermediate and high values of $\xi$, where the QE contributions
can be significant or even dominate the cross section, the indications of
scaling seen in previous data \cite{ne3_x} are confirmed.

\begin{figure}[xiscale]
\epsfxsize 8.5 cm \epsfysize 6 cm \epsfbox{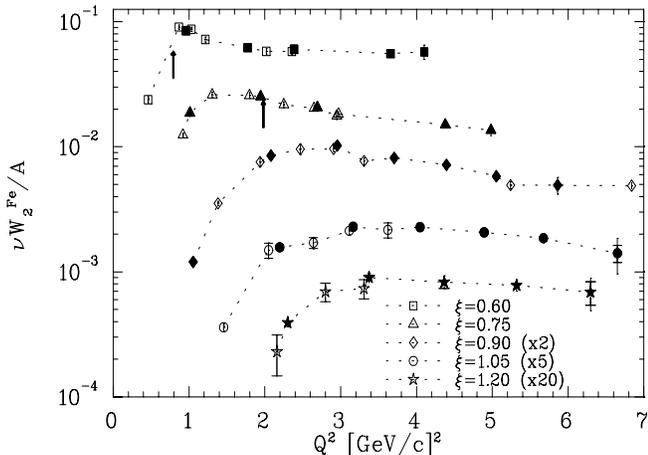}
\caption{Structure Function per nucleon for Fe as a function of $Q^2$. The
hollow points are from the SLAC measurements \protect\cite{ne3_x,ne18}. 
Dotted lines connect data sets at fixed values of $\xi$. The inner errors
shown are statistical, and the outer errors are the total uncertainties.  The
arrows indicate the position of the QE peak ($x=1$) for $\xi$=0.6 and 0.75.}
\label{xiscale}
\end{figure}

Figure \ref{xiscale} shows the structure function versus $Q^2$ at several
values of $\xi$.  At low values of $\xi$, we see a rise in the structure
function at low $Q^2$, corresponding to the QE scattering (at fixed $\xi$, low
values of $Q^2$ correspond to larger values of $x$). This is followed by a
fall to the high-$Q^2$ limit as the inelastic contributions dominate the
scattering.  Higher values of $\xi$, corresponding to $x$$>$1 for all $Q^2$ values
measured, contain
significant QE contributions. For all values of
$\xi$, the structure function is nearly constant, with variations typically
less than 10-20\%, for $Q^2 > 2-3$(GeV/c)$^2$.  Based on structure function
evolution observed at high $Q^2$ for fixed (large) values of $\xi$, QCD
scaling violations would be expected to cause roughly a 10\% decrease in $\nu
W_2$ for a factor of two increase in $Q^2$.

The new data allow a more careful test of the suggestion that the scaling
comes from an accidental cancellation of $Q^2$ dependent terms
\cite{accidental}.  In the PWIA, it was shown \cite{west,day} that the
quasielastic portion of the scattering should show scaling in $y$ (where $y$
is the minimum allowed momentum of the struck nucleon along the direction of
the virtual photon). Final-state interactions (FSI), neglected in the PWIA,
can cause scaling violations and introduce a $Q^2$ dependence to the quasielastic
scaling function, $F(y)$.

For very large values of $Q^2$ ($Q^2 >> M_N$), $y$ can be written in terms of
$\xi$, with corrections of order $1/Q^2$:

\begin{equation}
\label{accidental}
F(y) = F(y(\xi,Q^2)) = F( y_0(\xi) - \frac{M^3_N \xi}{Q^2} + O(1/Q^4 )),
\end{equation}
where $y_0(\xi) \equiv M_N(1-\xi)$.  The authors of ref. \cite{accidental}
argued that these $1/Q^2$ corrections would introduce scaling violations that
would be cancelled by final-state interactions.  Their nuclear matter calculations
indicated approximate cancellation of the $Q^2$ dependent terms from the FSI
and the variable transformation, leading to an accidental scaling of the
structure function in terms of $\xi$.  This cancellation occurs only for
intermediate $Q^2$ values, and it was predicted that the scaling would break
down at very large momentum transfers.
We can test this model by directly examining the size of the scaling
violations coming from FSI and from the variable transformation. The
violations coming from the transformation from $y$ to $\xi$ can be very large.
 At $y=-0.3$ GeV/c, which corresponds to $\xi \approx 1.1$ for $Q^2>2$
(GeV/c)$^2$, the scaling violations from the exact transformation are 
$>$200\% between $Q^2$= 2 (GeV/c)$^2$ and $Q^2$= 4 (GeV/c)$^2$, and $\gtorder$$50$\%
between $Q^2$= 4 (GeV/c)$^2$ and $Q^2$= 6 (GeV/c)$^2$.  In order to see
scaling in $\xi$, these large scaling violations would have to be cancelled by
FSI.  A $y$-scaling analysis of the new data \cite{89008_y} indicates that
final-state interactions produce $\ltorder$ 10\% deviations from scaling for
these values of momentum transfer, far too small to cancel the transformation
induced scaling violations.  

While the data show that the cancellation suggested in ref. \cite{accidental}
does not explain the observed scaling, the quality of the scaling indicates
that there is some connection between the $y$-scaling picture of quasielastic
scattering and the $\xi$-scaling picture of the DIS.  The $\xi$-scaling
analysis involves removing only the Mott cross section, while the $y$-scaling
analysis also removes the strongly $Q^2$-dependent elastic form factor, yet
both show scaling above $Q^2 \gtorder 3$(GeV/c)$^2$ in the region of low
energy loss.  In this region, the cross section is dominated by quasielastic
scattering and there is no expectation that $\xi$-scaling should be valid. 
While the connection between $\xi$-scaling and $y$-scaling in nuclei is not
fully understood, it is essentially the same behavior as seen by Bloom and
Gilman \cite{bgdual} in resonance scattering from a free proton. They measured
$\nu W_2^p$  as a function of an improved scaling variable, $x^\prime =
Q^2/(2M\nu+M^2)$, and observed that while there was significant resonance
scattering at high $x^\prime$ and low $Q^2$, the resonance structure, when
averaged over a range in $x^\prime$, agreed with the DIS limit of the
structure function.   The resonance peaks fall more rapidly with $Q^2$ than
the DIS contributions, but at the same time move to larger values of
$x^\prime$. The DIS structure function falls with increasing $x^\prime$, at a
rate which almost exactly matches the falloff with $Q^2$ of the resonance (and
elastic) form factors. This behavior also holds when examining the structure
function in terms of $\xi$ instead of $x^\prime$ \cite{niculescu,duality1}
(note that in the Bjorken limit, $x = x^\prime = \xi$).

\begin{figure}[xid]
\epsfxsize 8.5 cm \epsfysize 6 cm \epsfbox{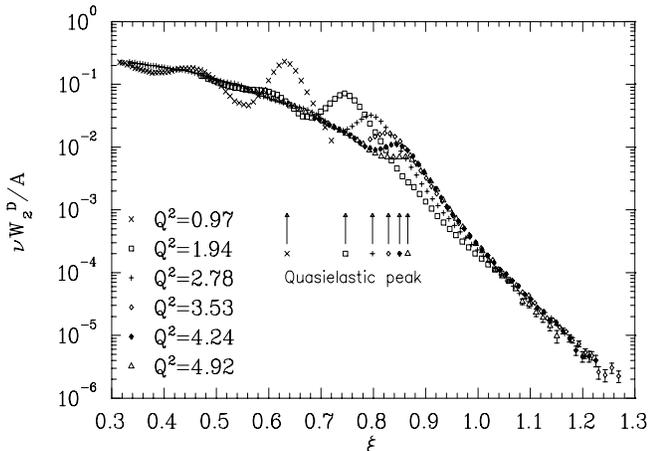}
\caption{Structure Function per nucleon for deuterium. The $Q^2$ values are
given for $x = 1$.  Statistical errors are shown.}
\label{xid}
\end{figure}

In nuclei, this same behavior leads to scaling in $\xi$.  When $\nu W_2^A$ is
taken as a function of $\xi$, the QE peak falls faster with $Q^2$ than the
deep inelastic scattering component, but also moves to larger values of $\xi$. In the
case of the proton, the resonance behavior follows the scaling limit on
average, but the individual peaks are still visible.  In heavy nuclei, the
smearing of the peaks due to the Fermi motion of the nucleon washes out the
individual resonance and quasielastic peaks, leading to scaling at all values
of $\xi$. Figure \ref{xid} shows the structure function versus $\xi$ for the
deuteron. Because of the smaller Fermi motion in deuterium, the QE peak is
still visible for all values of $Q^2$ measured and the scaling seen in iron is
not seen in Deuterium near $x$=1 (indicated by the arrows in Fig \ref{xid}).
Note that for $Q^2 \gtorder 3$(GeV/c)$^2$, the data still show scaling in
$\xi$ away from the QE peak.

The success of $\xi$-scaling beyond the deep inelastic region opens up an
interesting possibility.  In the Bjorken limit, the parton model predicts that
the structure functions will scale, and that the scaling curves are directly
related to the quark distributions.  At finite (but large) $\nu$ and $Q^2$,
scaling is observed and it is therefore assumed that the structure functions
are sensitive to the quark distributions.  It is not clear that this assumption
must be correct, but the success of scaling is taken as a strong indication
that it is true.  In nuclei, we see a continuation of the DIS scaling even
where the resonance strength is a significant contribution to the structure
function.   This opens up the possibility of measuring quark distributions in
nuclei at lower $Q^2$ or higher $x$.  If one requires that measurements be in
the deep inelastic regime (typically defined as $W^2 > 4$(GeV/c)$^2$, where
$W^2$ is the invariant mass squared of the final hadron state), data at large
values of $x$ can only be taken at extremely high values of $Q^2$. Because the
quark distributions become small at large $x$, and the cross section drops
rapidly with $Q^2$, it can be very difficult to make these high-$x$
measurements in the DIS region.  However, the observation of $\xi$-scaling
indicates that one might be able to use measurements at moderate values of
$Q^2$, where the contributions of the resonances are relatively small compared
to the DIS contributions and where these contributions have the same behavior
(on average) as the DIS.

A more complete understanding of $\xi$-scaling, through precision measurements
of scaling in nuclei and local duality in the proton is required.  High
precision measurements of duality in the proton have been made recently at
Jefferson Lab \cite{niculescu,duality1,duality3}, and additional proposals
have been approved that will extend these measurements to higher $Q^2$
\cite{dualprops}.  There is also an approved experiment to continue $x > 1$
measurements at higher beam energies, which will extend the present study
of $\xi$-scaling in nuclear structure functions to significantly higher $Q^2$
\cite{extension}.  Finally, there is an approved experiment that will make a
precision measurement of the structure function in nuclei as part of a
measurement of the EMC effect \cite{E00101}, which will make a quantitative
determination of how far one can extend scaling in nuclei when trying to
extract high $x$ nuclear structure.

In conclusion, we have measured nuclear structure functions for $x \gtorder 1$
up to $Q^2 \approx 7$(GeV/c)$^2$.  The cross section for $x > 1$ is dominated
by quasielastic scattering and, as expected, does not exhibit the $x$-scaling
predicted for parton scattering at large $Q^2$.  However the data do show
scaling in $\xi$, hinted at in previous measurements.  The $\xi$-scaling in
nuclei at large $x$ can be interpreted in terms of local duality of the
nucleon structure function, with nucleon motion averaging over
the resonances. Measurements of $\xi$-scaling and local duality,
combined with a more complete understanding of the theoretical underpinnings
of duality and $\xi$-scaling may allow us to exploit this scaling to access
high-$x$ nuclear structure functions, which can be difficult to obtain in the DIS limit.

We gratefully acknowledge the staff and management of Jefferson Laboratory for
their efforts. This research was supported by the National Science Foundation,
the Department of Energy and the Swiss National Science Foundation.



\begin{references}

\bibitem {georgi} H. Georgi and H. D. Politzer, Phys. Rev. D 14, 1829 (1976). 

\bibitem{ne3_x} B. W. Filippone \etal, Phys. Rev. C 45, 1582 (1992).

\bibitem{bgdual} E. Bloom and F. Gilman, Phys. Rev. D 4, 2901 (1971). 

\bibitem{derujula} A. DeRujula, H. Georgi and H. D. Politzer, Ann. Phys. 103, 315 (1977). 

\bibitem {accidental} O. Benhar and S. Liuti, Phys. Lett. B 358, 173 (1995). 

\bibitem{ne18} J. Arrington \etal, Phys. Rev. C 53, 2248 (1996).

\bibitem{89008_y} J. Arrington \etal, Phys. Rev. Lett. 82, 2056 (1999).

\bibitem{jra} J. Arrington, Ph.D. Thesis, California Institute of

Technology (1998).  URL address:

http://www.krl.caltech.edu/$\sim$johna/thesis.

\bibitem {bosted} P. Bosted \etal, Phys. Rev. C 46, 2505 (1992). 

\bibitem {dasu1} S. Dasu \etal, Phys. Rev. Lett. 60, 2591 (1988). 

\bibitem {dasu2} S. Dasu \etal, Phys. Rev. Lett. 61, 1061 (1988). 

\bibitem {dasu3} S. Dasu \etal, Phys. Rev. D 49, 5641 (1994). 

\bibitem {west} G. B. West, Phys. Rep. 18, 263 (1975).

\bibitem {day} D. B. Day \etal, Annu. Rev. Nucl. Part. Sci. 40, 357 (1990). 

\bibitem {niculescu} I. Niculescu, Ph.D. Thesis, Hampton University (1999).

\bibitem {duality1} I. Niculescu \etal, Phys. Rev. Lett. 85, 1186 (2000).

\bibitem {duality3} R. Ent \etal, to be published, Phys. Rev. D (2000).

\bibitem {dualprops} C. E. Keppel \etal, JLab Experiment E97-010,
(1997); I. Niculescu \etal, JLab Experiment E00-002 (2000).

\bibitem {extension} A. Lung, D. Day, and B. Filippone, JLab Experiment 
E99-015, (1999).

\bibitem {E00101} J. Arrington, JLab Experiment E00-101, (2000).

\end{references}
\end{document}